# Quantum Computers, Predictability, and Free Will

Gil Kalai

**Abstract:** This article focuses on the connection between the possibility of quantum computers, the predictability of complex quantum systems in nature, and the issue of free will. The author's theory, that quantum computation is impossible (in principle) directly implies that the future of complex quantum systems in nature cannot be predicted. A more involved argument shows that the impossibility of quantum computation supports the view that the laws of nature do not in fact contradict free will. At the center of the argument is the ambiguity inherent in the way the future is determined by the past; ambiguity that is not in terms of the mathematical laws of physics (which are fully deterministic) but rather in terms of the physical description of the objects we refer to.

## 1. Preface

My purpose in this article is to present a philosophical path, starting with the question of whether quantum computers are possible, moving on to the claim that it is impossible to predict the future of complex quantum systems in nature, and concluding by supporting the philosophical approach that sees no contradiction between free choice and the laws of nature.

Let us describe the three central topics of this paper. The first topic, quantum computers, deals with a new type of computer based on quantum physics. When it comes to certain computational objectives, the computational ability of quantum computers is tens, and even hundreds of orders of magnitude faster than that of the digital computers we are familiar with, and their construction will enable us to break most of the current cryptosystems. While quantum computers represent a future technology, which captivates the hearts and imaginations of many, there is also an ongoing dispute over the very possibility of their existence. The second topic is the question of predictability, which concerns the ability to predict the future of a complex system. This system can be a molecule, the weather, or a living being. In certain cases, such as the movement of planets (in our solar system) or the computational process of a reliable computer, we are fully capable – or rather almost fully capable – of predicting the future. The question is whether this possibility even exists, at least in principle, for every physical system, and specifically for a person's decisions. The third topic is the issue of free will. The concept of free will states that one is capable of making decisions based on free will, which will accordingly shape one's future. This is the way most people perceive their decisions, and hence they consider their actions as the outcome of a choice that is not determined by other factors. The question of free will originates in the apparent contradiction between the determinacy of the decision, stemming from the fact that the laws of nature determine the future by the past, and the above-mentioned sense of freedom. A common definition of free will is based on two requirements. The first is that there are several possibilities for the future, the possibility that will occur is not determined by the past, and, similarly, even events that have already occurred could have transpired differently from the way they actually did. The second requirement is that there is a component that concerns the question of which future will take place, and that solely depends on a person's decisions. Hard determinism, which negates the existence of free will, recognizes the causal results of our actions, yet denies our freedom of choice.

The link between a scientific debate on the issue of quantum computers, which has been the focus of our attention for several decades, and questions of predictability and free will, having their origin in ancient philosophy, might seem far-fetched at first. However, issues of computers and computation are naturally related to questions of predictability and free will, and from the dawn of modern interest in computation there have been several attempts to understand the issue of free will vis-à-vis this theory. The attempt to link the issue of free will to quantum physics was also set in motion shortly after quantum physics was born. Certain people regarded the randomness of the results of quantum physics measurements as a possible basis for unravelling the underlying tension between determinism and free will. Others, among them Schrödinger – a pioneer of quantum theory – absolutely rule out this possibility (Schrödinger 1936). For a thorough discussion with historical perspective of quantum physics and free will the reader is referred to Hodgson (2002).[1]

In this paper, as in several other papers that relate predictability and free will to quantum mechanics (see, e.g., Aaronson 2016), the notions of determinism and predictability also extend to situations where the future is determined or can be predicted probabilistically. For us, the term "unpredictability" refers to situations where the future cannot be predicted, not even probabilistically; "indeterminism" refers to situations where the future is not determined, not even probabilistically, from the past; and "multiple possibilities for the future" refers to situations where there are multiple probabilistic possibilities for the future, rather than a single probability distribution describing it. (See Figure 1.)

---

[1] For recent papers on the connection between free will and quantum physics see Briegel (2012), Aharonov, Cohen, and Shushi (2016), and Aaronson (2016).

Following is a brief description of the structure of this paper. In Section 2 we discuss classical computation, in Section 3 we present quantum computers and in Section 4 we describe the author's theory that quantum computation is not possible. For a more detailed discussion of the subject see Kalai (2019, 2021). Later in the paper we discuss in parallel the "Sycamore" quantum computer of 12 computational units (qubits), and the human-being Alice, whose free will we attempt to analyze. In Section 5 we discuss predictability and explain why the theory of the impossibility of quantum computers implies that complex quantum systems in nature, including a person's future decisions, are in fact unpredictable. This argument of unpredictability applies even to probabilistic prediction. There is still a great distance between the epistemic argument that a physical system is unpredictable and the ontological argument for multiple possibilities in the future, and we will discuss this in Sections 6 and 7. Section 8 will discuss the question of how multiple choices in the future enable the existence of free will. In Section 9 we link the discussion to reductionism and emergent theories, while Section 10 presents a summary of the entire argument.

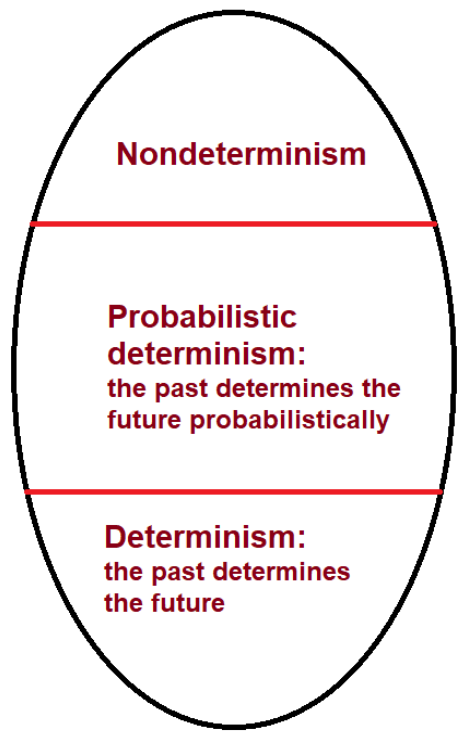

Figure 1: In this paper, the notions of determinism and predictability extend also to situations where the future is determined or can be predicted probabilistically.

## 2. Classical Computers

The classical computer can be seen as a device that contains n "bits" where every bit is in one of two positions of either "zero" or "one." The computer operates through "gates" that perform logical operations on either a single or two bits. Two simple types of gates are sufficient to describe all forms of computation: the NOT gate, which reverses the value of a single bit, and the AND gate, which receives two bits as input and produces "one" if and only if the value of each of the bits in the input is "one." The model presented here for computers is called the Boolean circuit model and is close to the 1940s models of Church, Turing, and others, which preceded the advent of the digital computer and the computer revolution in the second half of the twentieth century.

In the 1960s and 1970s, computer science researchers came to an important insight regarding the inherent limitations of computers. There are certain computational tasks that are very easy to formulate (and even to check whether a suggested answer to them is correct), but which are nonetheless impossible to perform because they require too many computational steps. For example, if the number of computational steps for a problem described by an input of *n* bits is *$2^n$*, then this computation will not be feasible for the most powerful computers, even when n = 100. To describe feasible (or "efficient") computations, that is, computations that can actually be performed, an important assumption was added, according to which the number of computational steps (i.e., the number of gates) does not exceed a constant power (e.g., $n^3$) in the number *n* of bits describing the input. This definition represents an important step in the theory of computational complexity, while providing the central tool for the analysis of the computational power of computational models, algorithms, and physical computational systems.

The main lens through which we analyze the difficulty of computation is the asymptotic lens. For most computational tasks we cannot hope for a full understanding of the number of required computational steps as a function of the number of bits of the input, but we can gain understanding (at times rough and at times just conjectural[2]) of the computational difficulty by studying how the number of computational steps asymptotically depends on the input size. Experience gained in recent decades indicates that asymptotic analysis provides a good understanding of the practical difficulty of computational tasks.

---

[2] A central conjecture of the theory of computational complexity, **NP ≠ P,** asserts that not for every claim that has a short proof (which can be tested efficiently), there exists an efficient algorithm for finding a proof.

Asymptotic analysis is also important for computational processes in nature. The "strong Church–Turing thesis" (Pitowsky 1990) asserts that any computation in nature can be described as an efficient computation of a classical computer. (In this case we rely on the asymptotic definition of efficient computation.) Pitowsky's paper also explains some interesting philosophical connections of the strong Church–Turing thesis and connections with quantum computers, that we will discuss later.

We will now enrich the picture by adding probability. A single bit has two possible values, "zero" and "one," and it is possible to extend the Boolean circle model by allowing the bit to have the value "zero" with probability *p* and the value "one" with probability *1-p*. In this way, the classical computer equipped with probabilistic bits can describe a sample from a probability space of sequences of zeros and ones. The model of probabilistic classical computers is an important model in the theory of computational complexity and it also constitutes a convenient introduction to the concept of quantum computers. Probabilistic Boolean circuits can be realized rather easily by digital computers.

## 3. Quantum Computers

The model of quantum computers is a computational model based on quantum mechanics, and was first introduced in the 1980s. In quantum computers, the classical bit is replaced by the basic computational element called a qubit. The quantum computer has *n* qubits. (A full description of the quantum computer model is complex, and even our partial and simplified description in the next two paragraphs may use notions that are not familiar to some readers. The reader can skip these paragraphs and rely only on the description of the Sycamore quantum computer below.)

The state of a single qubit is described by a unit vector in a complex two-dimensional vector space (namely, it is described by four real numbers whose sum of squares equals 1). The uncertainty principle asserts that we cannot identify the precise state of the qubit, but rather we can measure it and obtain a probabilistic bit. One way to think about it is as follows: there are two basic states for the qubit and we denote them by $|0\rangle$ and $|1\rangle$, while the general state of the qubit is a "superposition" of these two basic states, namely, the general state of a qubit is a linear combination of the form

$$(a + bi)|0\rangle + (c + di)|1\rangle,$$

where $(a + bi)$ and $(c + di)$ are complex numbers $(i = \sqrt{-1})$ satisfying

$$|a + bi|^2 + |c + di|^2 = a^2 + b^2 + c^2 + d^2 = 1.$$

The state of the qubit can be measured and when a qubit in this state is measured, we obtain a probabilistic bit with a value 0 with probability $a^2 + b^2$ and 1 with probability $c^2 + d^2$. One possible physical realization of the two basic states of a single qubit is with the two energy levels of a Hydrogen atom, with quantum physics allowing the atom to be in a superposition of these basic states as described above.

The computer acts on *n* qubits through "quantum gates," which are the basic quantum computation operations. (The gates are described mathematically by "unitary operators.") At the end of the computation process, the state of the computer is measured, and this yields a single sample from a probability distribution of 0-1 strings of length n.  In this case too, the assumption is that for each efficient computation the number of gates is at most polynomial in *n*. The crucial fact about quantum computers is that they would allow us to efficiently achieve samples that cannot be efficiently achieved by classical computers.

In the 1990s, Peter Shor discovered that quantum computers would allow the execution of certain computational tasks hundreds of orders of magnitude faster than regular computers and would enable us to break most current cryptosystems. Under common computational complexity assumptions, efficient computations in quantum computers, including Shor's algorithm, would violate the strong Church–Turing thesis mentioned in the previous section[3].

It was around that time that early doubts concerning the model also surfaced: quantum systems are by nature "noisy" and unstable. The term "noise" refers to a deviation of the computer from the planned program, and in the case of a quantum computer any unwanted interaction or leakage of information leads to such a deviation. Therefore, in order to reduce the noise level, quantum computers must be isolated and protected from their environment. The key to a possible solution of the noise problem is quantum error-correcting codes, which will enable noisy quantum computers to perform the same computations conducted by the abstract noiseless model, provided that the level of noise can be reduced to below a certain threshold denoted by **α**.

It is a common opinion that the construction of quantum computers is possible, that the remaining challenge is mostly engineering-related, that such computers will be constructed in the next few decades, and that error-correcting quantum codes of the required quality will be developed in labs in the years to come. My standpoint, which will be presented in the next section, is that it will be fundamentally impossible to reduce the noise level to below the required threshold. As a result, it will be impossible to develop the quantum codes required for quantum computation, nor will it be possible to reach the target of "quantum computation supremacy," whereby a quantum computer performs computation that is extremely hard or even impossible for a classical computer.

---

[3] In his 1990 paper, Pitowsky asked whether quantum computers would make it possible to solve general problems in **NP**. The prevailing hypothesis today is that even quantum computers will not allow this.

**The Sycamore Computer**

The Sycamore computer, which serves as a key example in this paper, was constructed by scientists at Google (Arute et al., 2019). It acts on n qubits (n lies between 12 and 53) and performs several rounds of computation. The qubits are organized in a two-dimensional array, and in each round of computation a certain physical operation is carried out in parallel on individual qubits or on pairs of qubits. Once the computation is complete, the state of the computer is measured and we thus get a sample of length n, namely, a sequence of length n of zeros and ones. This procedure is repeated numerous times (in just 300 seconds) yielding hundreds of thousands of samples. The experiment conducted by the Sycamore computer for n=53 led to a declaration that "quantum supremacy" had been achieved, referring to a demonstration of quantum computation that is very hard or virtually impossible to obtain on a classical computer (this achievement is still very much in dispute, see the Appendix; if verified it is in tension with the argument I present in the next section). In the present paper we will discuss only the simplest case of the Sycamore computer where *n = 12*.

## 4. The Argument against Quantum Computers

My argument for the impossibility of quantum computers lies within the scope of quantum mechanics and does not deviate from its principles. In essence, the argument is based on computational complexity and its interpretation, and it is discussed in-depth in my papers (Kalai 2019, 2021). Also included in these papers is a discussion of general conclusions that derive from my argument and relate to quantum physics, alongside suggestions of general laws of nature that express the impossibility of quantum computation.

My argument mostly deals with understanding quantum computers on the intermediate scale (known as NISQ computers, an abbreviation of Noisy Intermediate Scale Quantum), that is, quantum computers of up to at most several hundreds of qubits. It is expected that on this scale we will be able to construct quantum codes of a quality sufficient for the construction of bigger quantum computers. It is further expected that on this scale the quantum computer will achieve computations far beyond the ability of powerful classical computers, that is, will achieve quantum computational supremacy. The Sycamore computer is an example of a noisy intermediate-scale quantum computer.

As specified in this section, it is my argument that NISQ computers cannot be controlled. Hence:

a) Such systems cannot demonstrate significant quantum computational advantage.
b) Such systems cannot be used for the creation of quantum error-correcting codes.
c) Such systems lead to non-stationary and even chaotic distributions.

Regarding the first item, we remind the reader that computational complexity theory provides tools for studying the computational power of models and physical computational devices. The reason NISQ computers cannot support quantum supremacy is that when we use computational complexity tools to understand the computational power of NISQ computers, we discover that they describe a very low-level computational class. This low-level computational class does not allow for any complicated computations, much less computational supremacy. My analysis draws computational conclusions for NISQ computers based on their mathematical model's asymptotic behavior.

Regarding the second item, the reason it is impossible to build quantum error-correcting codes is that it requires an even lower noise level than that required for demonstrating quantum supremacy. The meaning of the infeasibility of quantum error-correcting codes is that even a quantum computer operating on a single qubit is inherently noisy. It is to be noted that the argument that the noise level required for error-correcting codes is lower than the level required for quantum supremacy is generally accepted by both theoreticians and experimental physicists.

**Randomness, Noise, and Chaos**

We will now discuss the third item of the argument, which deals with the inherent chaotic behavior of quantum computers on the intermediate scale. The computational task, through which it is hoped that NISQ computers (for instance, the Sycamore computer) will achieve quantum supremacy, is a sampling task: the computer produces samples according to a specific distribution. The fact that the result of the experiment is probabilistic is a fundamental insight of quantum mechanics. Furthermore, the fact that a computer on this scale is noisy stems from the mathematical model, when the quality of the gates performed by the computer is taken into account.

Analysis of the intermediate-scale quantum computer model points to chaotic behavior across a broad range of noise levels. Here we use the term "chaotic" in the following strong sense: it will not be possible, even probabilistically, to predict the behavior of the computer, namely the distribution of the samples it produces. This chaotic behavior is caused by high amount of noise sensitivity of the quantum computer samples, namely, small fluctuations in the parameters describing the noise will have a big effect on the distribution of the samples produced by the computer.

The third part of my argument asserts that the samples of the Sycamore computer operating on 12 qubits will be chaotic, and that this is an inherent feature of samples produced by any quantum computer that will ever be built for the purpose of performing the same quantum sampling program. The combination of the second and third parts of my argument leads to the conclusion that a certain level of chaos (which cannot be lowered beneath a certain threshold) will necessarily also be present in a quantum computer with a single qubit.

The three components of my argument concerning the infeasibility of quantum computers, namely, the primitive computational ability of intermediate-scale quantum systems, the infeasibility of good-quality error-correcting quantum codes, and the chaotic nature of quantum computer samples, will be put to the test in the extensive worldwide experimental efforts to build intermediate-scale quantum computers.

**A Brief Look at the Mathematical Analysis: The Fourier Expansion**

Following is a brief look at a central technical analytic tool that applies to all three components of the argument, namely, the Fourier expansion of functions. In our case we consider functions, whose values are real numbers that are described on sequences of length *n* of zeros and ones, and every such function can be written (as a linear combination) by means of a special set of functions, known as Fourier–Walsh functions. Fourier–Walsh functions can be sorted according to their degree, which is a natural number between 0 and *n*. Much as the regular Fourier expansion makes it possible to describe a musical sound as a combination of pure high and low tones, in Fourier–Walsh functions, too, the degrees can be seen as analogous to the heights of the pure tones.

Our first step is to study the Walsh–Fourier transform of the probability distribution of the 0-1 sequences in an ideal noiseless quantum computer, and the second step is to study the effect of the noise. The main technical component of my argument is that the noise will exponentially lower the coefficients of the higher-degree Fourier–Walsh functions. This leads to a mathematical distinction (Benjamini, Kalai, and Schramm 1999) between distributions that can be described by low-degree Fourier coefficients and are referred to as "noise stable" and distributions that are supported by high-degree Fourier coefficients and are referred to as "noise sensitive." A general distribution can have both noise-stable and noise-sensitive components. This is the reason that, on the one hand, the stable component in the noisy distribution is described by low Fourier–Walsh levels, and hence this component represents an extremely low computational power, and that, on the other hand, if the unstable part, namely, the contributions of the higher-degree Fourier–Walsh functions remain substantial, this contribution will be chaotic.

**On Classical Information and Classical Computation**

The question "why does this argument fail for classical computers?" is an appropriate critique of any argument asserting that quantum computers are not possible. Regarding my argument, the answer to this question is as follows: the path to large-scale quantum computers requires building quantum error-correcting codes on NISQ systems, and my theory asserts that this is impossible. At the same time, the primitive computational power represented by NISQ systems still allows for stable classical information, and subsequently even for classical computation.

**The Weak Link in My Argument**

The mathematical analysis that leads to the conclusion that noisy intermediate-scale quantum computers have primitive computational power is asymptotic. That is, a mathematical model of these computers is described and analyzed when the noise level is fixed, and the number of qubits increases. The conclusion I draw from this asymptotic behavior concerns the lowest noise level engineers can reach. I argue that it would be impossible to lower the noise level in such a way that enables us to create computers whose computational ability is low in an asymptotic analysis, but very high – indeed beyond the ability of classical computers – for several dozen qubits. The move from an asymptotic argument regarding the model's computational complexity to a concrete claim regarding engineering-related limitations of intermediate-scale computers is hardly standard, and my argument clashes with the strong intuition of experts in the field, according to whom an engineering effort would make it possible in principle (as well as in practice) to lower the noise level as much as we would like. Indeed, the multiple resources invested in building quantum computers, especially noisy intermediate-scale quantum computers, are based on the common position that there is no fundamental obstacle obstructing this effort.

**More on Chaos**

Let me briefly discuss the strong notion of chaos used in this paper. When we speak here of chaotic behavior we refer to a system (either deterministic, probabilistic, or quantum) that is so sensitive to its defining parameters that its behavior (or a large portion of its behavior) cannot be determined, not even probabilistically. Chaos in this sense is sometimes called "Knightian uncertainty" (Aaronson 2016), a term that originates in economics. Our notion of chaotic behavior is related to the mathematical theory of "noise sensitivity" (Benjamini, Kalai, and Schramm 1999) that we mentioned above, and to the related mathematical theory of "black noise" (Tsirelson and Vershik 1998). Both these theories have their early roots in Weiner's chaos expansion (Weiner 1938). Weiner's motivation came from chaotic phenomena in nature and his expansion (known later as "polynomial chaos") is used to determine the evolution of uncertainty in stochastic dynamical systems.

We note that the term "chaotic system" in mathematical chaos theory (Lorenz 1963) refers to nonlinear classical deterministic systems, whose development very much depends on their initial conditions and since these conditions are not known precisely, the system's development in the long run cannot be predicted. We get chaotic probabilistic systems (in our strong sense) from deterministic (classical) chaotic systems since classical deterministic chaotic systems will magnify even a small amount of noise sensitivity (or Knightian uncertainty) in their initial conditions (or in the description of their evolutions). It is plausible that this magnification of noise sensitivity is relevant for understanding natural chaotic systems like the weather, and that natural chaotic phenomena thus combine ingredients from both Lorenz's (well-known) theory of chaos and Weiner's (less-known) theory of chaos.

For a thorough discussion of chaos from a philosophical perspective, with connections to quantum mechanics, determinism, and free will, see Bishop (2017). For examples of chaotic behavior of quantum systems which might be relevant to our discussion see Panda and Benjamin (2021) and Berke et al. (2022).

## 5. On Predictability

For the rest of the paper, we discuss and compare the 12-qubit quantum computer "Sycamore", and the human-being Alice. In this section we study the question of predictability which is in itself one of the central topics of this paper.

**Hard Determinism and Hypothetical Computers**

It is possible to link the assumption of determinism and the question of predictability to the following claim:

In principle (perhaps even in practice), it is possible to enter the description of Alice's brain at a given time into the hypothetical computer **M**, which will allow us to fully predict Alice's actions at any later time, as a function of the signals received by Alice from the external world.

It is to be noted that as our universe is quantum, and as Alice's brain may consist of quantum processes that are relevant to her actions, the quantum computer model (which was discussed in Section 3) is particularly suitable to the description of our suggested hypothetical computer **M.** We will discuss **M**-type computers capable of computing the future of other quantum computers. In particular, we will focus on a computer **M(S)** capable of predicting the samples of the Sycamore quantum computer.

The question regarding the possible existence of an **M**-type computer has arisen in many contexts, including that of free will. The idea that an **M**-type computer is possible is indeed quite common, as is the claim that if an **M**-type computer is possible, then human free will is nonexistent.[4] Discussions on the "recording" of human consciousness to the computer, which would enable the computer to run backwards and forwards in time, are quite popular nowadays. There is an entire community – "the community of singularity" – focused on the possibility of recording and restoring instances of human consciousness, regarded by members of the community as a probable and indeed imminent reality. Our reply in this case is unequivocal: an **M**-type computer (or "weatherman") is not possible. One reason for this is that quantum computers far simpler than **M** are impossible, and the principles preventing their very existence also apply to **M.** Yet another crucial reason for the impossibility of **M** is related to chaotic behavior, which will be discussed below.

**The Impossibility of Predicting the Future of Complex Quantum Systems**

The impossibility of quantum computers leads to far-reaching conclusions (Kalai 2019, 2021). Let us first discuss quantum systems with a low noise level (entropy). Computationally speaking, these systems are very simple, as long as their behavior is stable and predictable. If they are somewhat complex, as in the case of the Sycamore quantum computer with 12 qubits, the systems' development over time is unpredictable. (As I mentioned in Section 4, even a quantum process of measuring a single qubit will necessarily lead to a chaotic probabilistic bit.) These conclusions apply to the human brain. It is reasonable to assume that the quantum component of processes in the brain is more complex than that of noisy quantum computers with a small number of qubits (and certainly more complex than a single-qubit quantum computer). Consequently, the quantum component, which adds chaotic randomness to the brain, makes it impossible to predict (or even reconstruct) the brain's development over time. That is, a computer of type **M**, which provides an approximate description of the development over time of intermediate-scale quantum systems, is impossible and this conclusion also applies to more complex quantum systems, such as the human brain.

---

[4] Scott Aaronson (2016) deals with the issue of free will from the perspective of computer science, quantum mechanics, and quantum computing. For Aaronson, the issue of the predictability of the human brain and the possibility of the **M**-type computer is a main part of the free will issue. Aaronson considered hypothetical objects called freebits that carry the seeds of unpredictability in the physical world. Sabine Hossenfelder (2012) also emphasizes predictability as central to the free will problem and offers a mathematical (toy) model of physics with freedom. John Conway and Simon Kochen (2006) identified free will with unpredictability and claimed that if humans have free will then so do elementary particles.

Let us now briefly discuss the Sycamore computer with 12 qubits. We claim that an **M**-type computer that predicts Sycamore's accurate sampling, henceforth referred to as **M$_1$(S)**, is by no means possible. Furthermore, even an **M**-type computer that predicts the distribution of the sampling, henceforth referred to as **M$_2$ (S)**, is not possible. One may well ask why the Sycamore computer cannot predict the distribution of its own samples. The answer is simple: the Sycamore computer's samples display nonstationary behavior (i.e., their empirical distribution changes significantly over time). The claim that the samples exhibit chaotic behavior is even stronger and implies not only that the Sycamore computer cannot be considered an **M$_2$(S)**-type "predictor" of its own sampling distribution, but that such a predictor is not possible at all.

Let us briefly elaborate on the brain. Our claim refers not only to overall brain activities but also to the question of the possibility of predicting decisions in a small segment of those activities, and in the latter case as well it applies not merely to the possibility of accurately predicting decisions, but also to the possibility of predicting the probabilities of the various possible decisions. Building a predictor of type **M** to predict the next moves of a human chess player (whether accurately or probabilistically) is probably harder than building a predictor of type **M$_2$(S)** for the Sycamore computer. (There is no difficulty in predicting the next move of a computer program in a chess game.) It should be noted that it is reasonable to assume that human intelligence in fact originates from brain processes that enable stable classical information, and that are reminiscent of a classical computer, while the quantum nature of brain processes adds to the computation an element of unpredictable randomness.[5]

A significant step is yet to be made between the epistemic claim that the future cannot be predicted and the ontological claim that there are various possibilities for the future. We will discuss this in later sections. We will now discuss the gap between ontology and epistemology in the context of predictability.

**Determinism and Predictability: Ontology versus Epistemology**

For those of us who are not familiar with philosophy (the author of this paper among them), we will mention the distinction between epistemology, the theory of knowledge, which deals with the question of what can be known about the world, and ontology, the theory of being, which deals with the essence of the world. The assertion that the future of a physical system cannot be predicted is an epistemic claim regarding what can be known about the world, and can be compared to the assertion that the past of a physical system cannot be known, for instance, the number of hairs on Napoleon's head at the time of his birth.

---

[5] Yet another prominent standpoint is that of Penrose (1989) who regards quantum processes in the brain as a basis for outstanding computational abilities.

Connecting assertions about the impossibility of predicting the future with free will might be seen by scholars of philosophy as an elementary mistake: if determinism negates the assertion that Alice has free will, it does not matter whether the deterministic future can be predicted or not. Despite this observation, there are reasons to regard the epistemological question of predicting Alice's future as one of central importance in the context of the free will issue. One reason is that when we attempt to discuss the free will problem as a scientific question and seek a scientific experiment that may support one of the opposing views, the practical possibility for predicting Alice's future becomes crucial. Another reason that is relevant to the connection with quantum theory is the fact that different interpretations of quantum mechanics lead to very different ontological descriptions of the world. We will discuss this briefly in the next section. (The significant gap between these ontological descriptions has no bearing on the outcomes of experiments in quantum physics.) My position is that a valid argument about the fundamental unpredictability of Alice's decisions is of significance for the free will issue, yet it does not suffice, and hence the challenge of examining the ontological situation must be met. We will attempt to do so in the following sections.

## 6. What is the Nature of the Sycamore Computer?

We suggest several options for describing the Sycamore computer built by Google engineers. According to hard determinism, the full and complete description of the Sycamore computer, including its complete interaction with the environment, makes it possible to fully determine its actions.

**Sycamore-1** is the complete physical description of the Sycamore computer, which includes the computation program along with a full description of every external signal that could affect its operation. The samples produced by Sycamore-1 are determined by the past.

**Sycamore-2** is Sycamore as described by quantum mechanics. The physical description of the quantum computer makes it possible to determine the *distribution* of the samples it will produce as a function of every external signal. The distribution of samples produced by Sycamore-2 is determined by the past.

**Sycamore-3** is the Sycamore computer as described in the previous section. The empirical distribution of the samples it produces is a chaotic variant of the distribution described by the quantum computer's program. Sycamore-3 samples cannot be predicted in advance, not even probabilistically.

It is possible to think that Sycamore-2 is but a partial and vague description of the Sycamore computer, which only approximates the complete description provided by Sycamore-1, yet from the standpoint of quantum theory, Sycamore-2 is the significant concept. Similarly, it is possible to think that even if, as claimed in the previous section, the distribution of samples of the Sycamore computer are chaotic and unpredictable, Sycamore-3 is still but a partial and vague description of the Sycamore computer, which only approximates the complete description provided by Sycamore-2. I claim that that our partial and vague understanding of Sycamore-3 is, as a matter of principle, the optimal understanding of the Sycamore computer and that the concepts of both Sycamore-2 and Sycamore-1 are, in fact, meaningless in the physical world.

The claim that it is not possible to predict the future of the Sycamore computer is an epistemic claim, but the claim in this section that a more accurate physical description of the computer that allows such a prediction does not exist is already an ontological claim, and it will lead to an ontological description of multiple possibilities for the future.

I argue that it is not possible to identify the Sycamore computer with a physical system for which the distribution of future samples is determined at present, because this requires moving to a physical system that is too large to be considered a physical description of the computer. Moreover (and this will be the topic of the next section), in such a large physical system the notion of time and that of the causal connection between past and future lose their meaning.

**Sycamore in Space: Aram Harrow's Hypothetical Quantum Computer**

The claim that in order to understand the samples of the Sycamore computer we must switch to a very large physical system, perhaps even the entire universe, is connected to a thought experiment suggested by Aram Harrow during a public scientific debate held with me concerning the feasibility of quantum computers (Kalai and Harrow 2012). Let us begin with the quantum computer (for instance, the Sycamore computer), which we will place in an isolated location between galaxies. In order to cancel the influence of noise – that is, interactions with the surrounding environment – on the quantum computer, we should simply include the environment in the very definition of the quantum computer. In quantum theory this procedure of adding the environment to the system is known as a "going to the Church of the larger Hilbert space," and it has to do with the fact, that, mathematically, a noisy quantum process is equivalent to a noiseless quantum process implemented on a larger space. My answer to Harrow's thought experiment was rather similar to the question of who will guard the guardians. That is, once we add the environment to the core definition of quantum computer, we should also add the environment affecting that environment, and so on ad infinitum.[6] Harrow eventually agreed that a repetitive inclusion of the "environment" in the definition of a quantum computer would necessitate a quantum computer inclusive of the entire universe. Despite our agreement on this particular point, our disagreement on the possibility of quantum computers still stands.

Aram Harrow's thought experiment is related to my argument that the concepts of Sycamore-2 and Sycamore-1 cannot have a physical description as a subsystem of the universe that is not the whole universe, and therefore in fact have no physical description at all. The question remains, if the future distribution of the samples of the Sycamore computer (and perhaps even the samples themselves) is determined from the description of the entire universe in the past, doesn't it suffice to support hard determinism?

**If one accepts the claim that the Sycamore computer has no physical description at present that determines the distribution of samples in the future, can this be seen as an ontological claim regarding multiple possibilities for the future?**

I tend to think that the answer is yes. If there is no physical description of the Sycamore computer that determines the distribution of its future samples, the meaning is that there are, ontologically, multiple possibilities for the future. These multiple possibilities for the future are derived from the inherent ambiguity of the physical description of the Sycamore computer. On the other hand, it can be argued that in principle the future distribution of samples is uniquely determined since the required information already exists in the present, even if this information is encoded in the entire universe. To address this argument, I will raise another complementary argument in the next section and claim that when we move to discuss the entire universe the distinction between past and future disappears and the causal fabric between past and future dissipates. This last claim is related to the interpretation of the mathematical nature of the physical laws of nature.

We conclude this section with a brief discussion related to the Sycamore computer, on the gap between different worldviews based on interpretations of quantum mechanics.

**What Does Quantum Mechanics Say: Sycamore-1 or Sycamore-2?**

---

[6] For a similar argument, see Hemmo and Shenker (2021).

We have defined Sycamore-1 as a complete (hypothetical) physical description of the Sycamore computer, which makes it possible to determine at present the samples that will be created by the Sycamore computer tomorrow. Sycamore-2 is the Sycamore computer as quantum mechanics describes it and it is a description that makes it possible to determine at present the distribution of samples that will be created by the Sycamore computer tomorrow. According to one interpretation of quantum mechanics, a complete identification of samples that will be created tomorrow has no physical meaning whatsoever, and therefore Sycamore-1 is devoid of meaning as well. If we adopt this interpretation, the correct ontological description of the future is probabilistic. Here, we take the position that an ontological view of the physical world cannot be based on physically meaningless notions.

According to another interpretation of quantum mechanics, the theoretical possibility of identifying the full behavior of the computer (i.e., the samples themselves and not just their distribution) exists but should be rejected as impractical. (The second interpretation is based on the fact that in reversible quantum processes, information is neither lost nor created and therefore the identity of the computer's samples tomorrow is already encoded in the universe today.) These different interpretations of quantum mechanics lead to a large gap in the ontological view of quantum reality. (For example, there is no uniform answer to the question of whether the future is probabilistic.) But it is generally accepted that this gap has no bearing on any scientific experiment in quantum physics.

Recall that my theory (which has far-reaching experimental consequences) claims that the description given by Sycamore-1 or by Sycamore-2 has no physical meaning. This leads to an ontological view that the future is inherently chaotic, namely, that there are multiple possibilities for the future even if we consider a probability distribution for various future events as a single possibility. These multiple possibilities for the future of the Sycamore computer are not consequence of the inability of Google's scientists who built the computer, but rather a fundamental feature of every 12-qubit quantum computer that will ever be built for performing the tasks of the 12-qubit Sycamore computer.

## 7. On the Redundancy of Mathematical Models and the Nature of Time

In this section I will explain (by means of an analogy) the alternative I propose to hard determinism. The key point is that our perceptions of the concept of time and the causal connection between past and the future are valid only in limited and noisy physical systems, while on the broader scale of the entire universe the concept of causality between past and future events and even the concept of time itself[7] (largely) lose their meaning.

In limited and noisy systems there is a causal connection between past and future, but because of the influence of interactions with all that is external to the system – an influence that indeed may be crucial – the future is not completely determined by the past. For many quantum systems, any attempt to expand the system so that it also includes all external factors of influence will inevitably end (much like Aram Harrow's thought experiment) with the universe. On this large scale, describing the system as a function of time does indeed express a significant redundancy of the mathematical model, but the concept of causality linking the past and the future loses its meaning. The meaning of "redundancy" in a mathematical system is that partial information about the system makes it possible to fully reconstruct the system. The fact that the future is determined by the past for the whole universe expresses the redundancy of the mathematical description of the universe and, in exactly the same way, for the whole universe, the past is determined by the future.[8]

The following simple example demonstrates the meaning of redundancy in a mathematical system. Figure 1 shows a board of squares, in which the squares are colored in red and blue. The rule is that in every 2x2 square, the number of red squares is even. The mathematical rule leads to a significant redundancy of the coloring implemented on the board. For instance, the squares in the bottom row and the left column could be colored arbitrarily, and this coloring singularly determines the coloring of the rest of the squares (similarly, knowing the colors of the third row from the bottom and the second column on the right allows us to determine the colors of all the squares). Our model has large mathematical redundancy (resulting from the coloring rule), but, nevertheless, there is no past, future, or causality here. For example, it is incorrect to say that the colors of the squares in the bottom row and the left column are the reason why the top right square is red. One can think of a process where initially, at $t = 0$, the colors of the squares in the bottom row and the left column are arbitrarily determined and then, at $t = 1$, the colors of the squares for which all the neighboring squares are already colored are determined, and so on. But this is just one description that gives meaning to "time" in our model, there are other

---

[7] The suggestions urging the need to view time as an expression of the increase of entropy in physical systems have already come up in the context of thermodynamics in classical physics. Lloyd (1988) argued that time is actually an expression of quantum noise (decoherence), which characterizes limited and noisy systems.

[8] A universal quantum computer makes it possible to reverse the arrow of time and to describe any quantum process implemented on any geometry. By contrast, for the noisy quantum computation model, at a noise level that does not enable quantum error-correction, the arrow of time cannot be reversed, and it is no longer possible to implement any quantum process on an arbitrary geometry. These observations suggest a link between the fundamental impossibility of correcting quantum noise and the emergence of geometry and time.

descriptions as well, and the model can be understood without any reference to the concept of time.

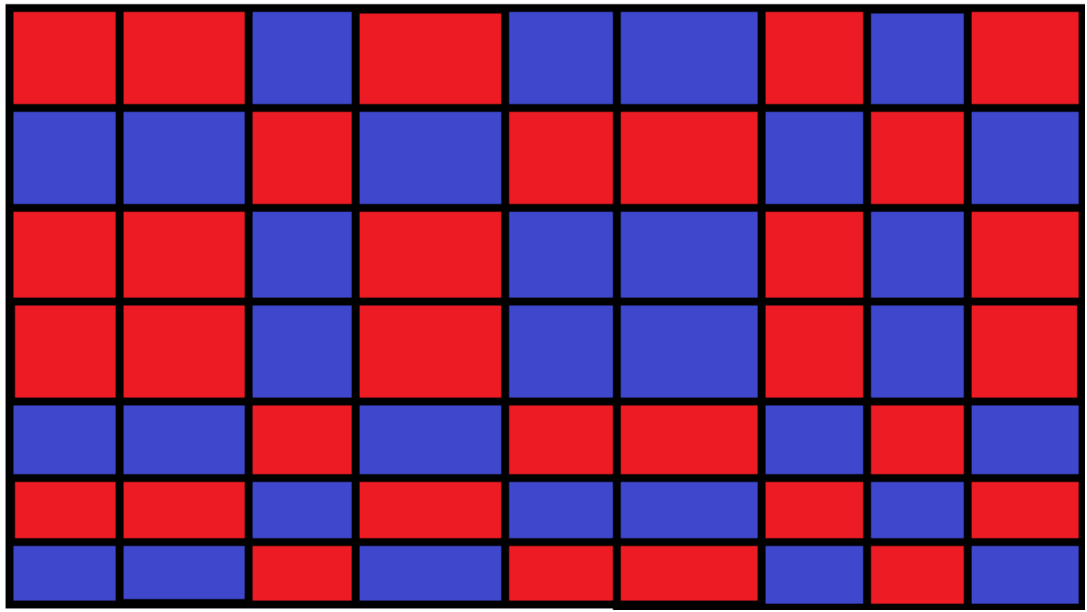

Figure 2: An example of redundancy in a mathematical system. A 7x9 board colored in such a way that in every 2x2 square there is an even number of red squares.

My argument about time is that a description of the universe as a system that evolves over time[9] expresses mathematical redundancy, similar to our simple checkerboard model. This redundancy does not necessarily express a causal connection between past and future.[10] The arrow of time and the concept of causality between past and future are expressed in partial and noisy systems, but in these systems, there are claims about future events that are not determined by the past. This argument makes it possible to offer an alternative to hard determinism, which is schematically described in Figure 2, in which there are claims about future events that are not part of the causal fabric between past and future. These claims include the identity of the next sample of the Sycamore computer, as well as an answer to whether Alice will eat ice cream tomorrow. These claims even include the next sample distribution of the Sycamore computer and the answer to the question of what the probability is that Alice will eat ice cream tomorrow. When it comes to the questions we have described, it is possible to say that in every physical system in which the concept of time and the causal connection between past events and future events maintain their meaning, there are also several possibilities for the future. There are also questions that are part of the causal fabric between past and future like those regarding the outcomes of a classic computer and the movements of planets in our solar system.

---

[9] Indeed, physicists have noted fundamental problems with the description of the universe as a system that expresses "evolution in time," and even with an exact mathematical expression of time. (The size of the universe can give an approximate description of time.) On the face of it, it seems that their view reinforces the picture we offer.

[10] Bertrand Russell (Russell 1912–1913) argued that the concept of causality is a primitive concept that is eliminated from any scientific description that uses differential equations to describe the laws of nature. See also Mark Steiner's article (Steiner 1984).

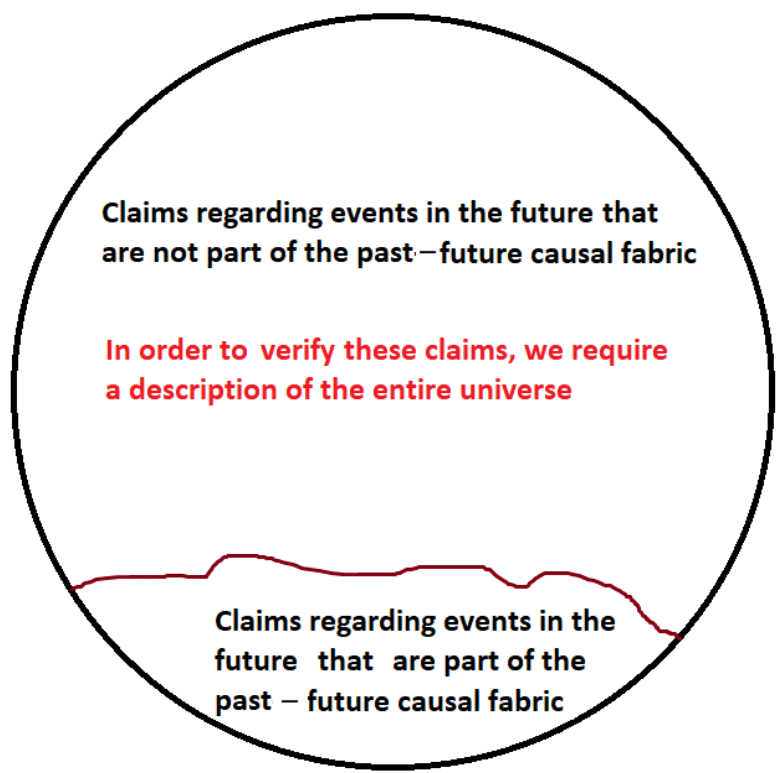

Figure 3: A schematic description of my proposed alternative to hard determinism.

Future events that are not part of the causal fabric between past and future can include events about the behavior of chaotic systems such as the weather. The weather in 2030 is described by a chaotic mathematical (classical) system, and so it is reasonable to think that quantum fluctuations that affect the initial conditions of the system and its evolution make it impossible to predict now in 2022 the weather in 2030, not even probabilistically. This claim about the weather is epistemic but, like for the Sycamore computer with 12 qubits, we offer here an ontological framework for understanding the multiplicity of possibilities regarding the future weather in 2030.

Having several possibilities for the future is an important ingredient for the concept of freedom, but an additional component is required: we must explain not only why there are several possibilities for the future, but also why certain components of Alice's future depend solely on her decisions. This is where Alice and the Sycamore computer part ways: there are various possibilities for the future of the Sycamore computer but, naturally, it does not have any will of its own. In the following section we will try to understand Alice's nature, much like the discussion in the previous section regarding the nature of the Sycamore computer.

## 8. On Alice's Free Will: Who Are You, Alice?

Alice wished to have fruit and hesitated between a nectarine and an apple that were both on the fruit tray. Having considered the taste of each fruit as she remembered it, the bother of cutting the apple and removing the seeds, and the fact that this was the very last nectarine on the tray, she chose the nectarine and ate it. Was Alice's choice an expression of her free will? If all the events we've described were causally determined by the state of the world in the dinosaur era (or during the Big Bang), then the free will perception might be nothing more than an illusion. In this section we will ask whether a deeper or more detailed description of Alice as a physical system will once again raise the possibility that Alice's future (and specifically her choices and desires) is predetermined or, if Alice's future is not predetermined, whether certain components in Alice's future are determined solely by her decisions.

According to hard determinism, insofar as Alice's actions affect her future, her actions are not free, but are rather causally derived from the past, and hence Alice's full and accurate description determines her future as a function of all the signals she will receive from her environment. Our concept of Alice (like Alice's concept of herself) is simply a partial description of her (this is indisputable), and our (or Alice's) perception that Alice is free to shape her future is nothing more than an illusion, stemming from our limited ability to understand Alice.

Let us once again describe the three concepts of Alice as they appear in the previous paragraphs, analogously to the descriptions of the Sycamore computer in Section 6.

**Alice-1** is the complete physical description of Alice, which includes every single external signal that may affect her future. Alice-1's future, including her actions, is determined by the past.[11]

**Alice-2** is the complete physical description of Alice, which makes it possible to <u>probabilistically</u> determine her future as a function of the past. Alice-2's future, including her actions, is probabilistically determined by the past.

**Alice-3** is Alice as she or any other person (even one equipped with any measuring device or computer) may describe her. Alice-3 is a partial, subjective, and vague description.

---

[11] It is also possible to make distinction between a physical description that includes the external signals and a physical description that is a *function* of the external signals. This distinction does not change our argument.

As previously mentioned, one may think that Alice-3 is but a partial and vague description of Alice that only approximates the complete description given by Alice-2 or even Alice-1. However, I would like to argue that the concepts of Alice-1 and Alice-2 are actually meaningless in the physical world, much like the examples of Sycamore-1 and Sycamore-2 in Section 6. In order to view Alice's future as part of the causal fabric deriving from the past, we must switch to a physical system so wide that the concept of time and the concept of causality between past and future lose their meaning. The only concept that has any physical meaning is that of Alice-3, despite it being partial and vague. When we identify Alice with Alice-3, Alice-3's future is not determined in the present, just as the future of much simpler noisy quantum systems are not determined in the present. Similarly, when we identify Alice with Alice-3, then Alice's decisions and actions in the present are not determined by the past. The perception of Alice-3 as the correct ontological concept of Alice (being the only one which has physical meaning), together with the ontological argument for multiple future possibilities made the in the previous sections, allows several possibilities for Alice's future, enables her decisions and actions in the present to affect her future, and refutes the claim that these very decisions and actions have already been determined in the past. While my argument does not verify the existence of freedom, it does enable its existence and negates the argument that freedom of choice contradicts the deterministic laws of physics.

I argue above that the concepts of Alice-1 and Alice-2 are actually meaningless in the physical word. Let us once again compare Alice to the Sycamore computer. As previously mentioned, although it is meaningful to claim that the first next-day series of ones and zeros produced by the Sycamore computer is 010010100111, this claim is not determined by the causal fabric between past and future, and it is indeed impossible to give physical meaning to the Sycamore computer today in a manner that determines the correctness or even just the probability for this future event. Similarly, we comprehend the meaning of the sentence "Alice will have a hard-boiled egg tomorrow," but it is not determined by the causal fabric between past and future, and it is impossible to give Alice a physical meaning today in a manner that determines the correctness or even just the probability of this claim.

**Alice as a Huge Set of Possibilities**

Another way in which we can view Alice-3 is as a huge set (or, to use a mathematical term, as a huge equivalence class) that includes all the hypothetical possibilities for Alice-1 (or Alice-2). Not only is it impossible to physically identify Alice as a specific member of this huge group, but the correct description of Alice is as the set itself and not as a specific member of it. It is possible to view this set as an expression (which changes over time) of "all that is not chaotic" in Alice.

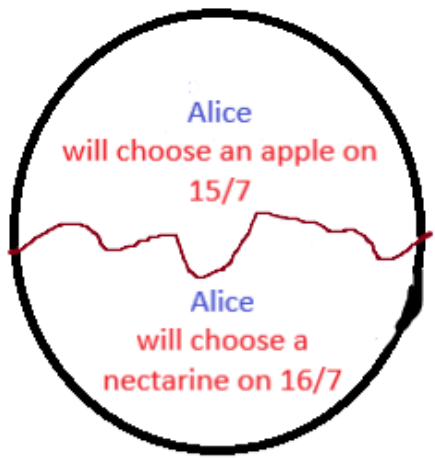

Figure 4: The set of possibilities describing Alice on 15.7. According to my proposed explanation, on July 15, it would be physically meaningless to differentiate between Alice's two possible choices on the following day, July 16.

The explanation I propose in this section concurs with that of Greek and later philosophers regarding free will as an expression of the small gap created by the lack of knowledge about causality. We can further add to this that the gap in question (which is not necessarily small) does not express vagueness in the mathematical description of the laws of nature but rather vagueness in the physical description of the objects we are discussing. Alice-3 can also be regarded as an emergent concept, and we will briefly elaborate on this in the next section.

## 9. Reductionism, Emergent Concepts and Another Look at Predictability

**The Perception of Science as a Layered Structure versus Reductionism**

Physics can be seen as a layered structure. Quantum mechanics provides a general framework for the laws of nature, while other layers offer a more specific description of the laws of physics in our world. There is, however, another approach, taken by quite a few scientists both explicitly and implicitly, which can be defined as "hard reductionism." According to this approach, there should be a method that, in principle, will allow every physical principle or physical theory to be deduced from quantum mechanics. (A popular variation of this approach is the claim that all laws of nature can be deduced from theories that deal with elementary particles, at the center of which lies the Standard Model). An interesting test case revolves around the question of whether the laws of thermodynamics can be derived from quantum mechanics (Hemmo and Shenker 2001). Reductionism, as opposed to a layered perception in science, is an important philosophical issue in and of itself (Putnam 1973, Ben-Menahem 2018), and has also been addressed by physicists (Anderson 1972). In physics, it is sometimes customary to distinguish between fundamental theories and emergent theories (also known as effective or phenomenological theories), with the latter providing a description of physical systems and phenomena without presuming to describe the reasons for the correctness of the description, reasons that must eventually be sought in fundamental theories. According to the layered perception, emergent theories, which provide the best picture *known to us* for understanding systems in physics and chemistry, give, in many cases and as a matter of principle, the best possible understanding of these systems. It is wrong to think that fundamental theories will provide a better understanding, and this in no way contradicts the correctness of more fundamental theories. The means through which fundamental and emergent theories "correspond" with each other is stable classical information (which function as basic constants of nature) that is transferred from more fundamental layers to higher emergent layers.

My theory, which negates the possibility of quantum computers, complies with the layered perception of physics. While it lies within quantum mechanics it also points to the fact that not everything that is theoretically enabled by quantum mechanics is necessarily supported by emergent higher-layer theories. The description of a world devoid of quantum computing concurs (in several respects – both conceptual and technical[12]) with a layered perception of physics itself and of science in general.

**Alice as an Emergent Concept and the Connection to Free Will**

Many have discussed the connection between the layered perception of reality and the issue of free will, and the possibility that the key to solving the problem is based on a higher-layer understanding of Alice and her choices.[13] It is especially worth mentioning Donald Davidson's theory (Davidson 1970); one of the underlying principles of Davidson's theory negates the possibility of accurately basing mental states on the laws of nature, further asserting that it is impossible to either explain or predict mental states on the basis of deterministic physical laws.

Our discussion in the previous section is in line with this principle. As I argue there, Alice's emergent description (Alice-3) has no alternative as a more basic or more accurate physical description, and this claim in turn is close to Davidson's assertions regarding mental states; what makes my approach unique is the claim that the principles of the emergent description apply to much larger physical systems (such as Sycamore-3), thus extending far beyond the context of the free will issue.

---

[12] For example, the lack of quantum error-correction leads to the fact that the noise in higher-layer physical theories mainly expresses the symmetry and laws of the high-layer theory, rather than the symmetry and laws of a lower-layer theory.

[13] In this context we wish to also mention Jenann Ismael's book (Ismael 2016) and Carlo Rovelli's paper (Rovelli 2013).

**If Alice's Future is Unpredictable, are Quantum Computers Necessarily Impossible?**

Our argument in Sections 5-8 is accordingly based on the claim (Section 4) that a Sycamore-type computer that operates on 12 qubits without (almost) any noise is not possible. The infeasibility of quantum computers is a drastic option, and we conclude from it that there are claims about the future that cannot be predicted and cannot be considered as part of the past–future causal fabric. This argument applies to the computer proposed by Aram Harrow as well as to the results of the Sycamore's first sampling (whether we're referring to the actual sampling or to the distribution describing it). The argument also applies to accurate forecasting of the weather, and to the question of whether Alice will have a hard-boiled egg tomorrow – regardless of whether we wish to have a deterministic or a probabilistic answer. The separation I am proposing contradicts hard determinism and the boundary I am suggesting negates the possibility of quantum computing (Figure 4). It is, however, possible to think of boundaries that do enable quantum computing, but I am not familiar with a theory that supports them.

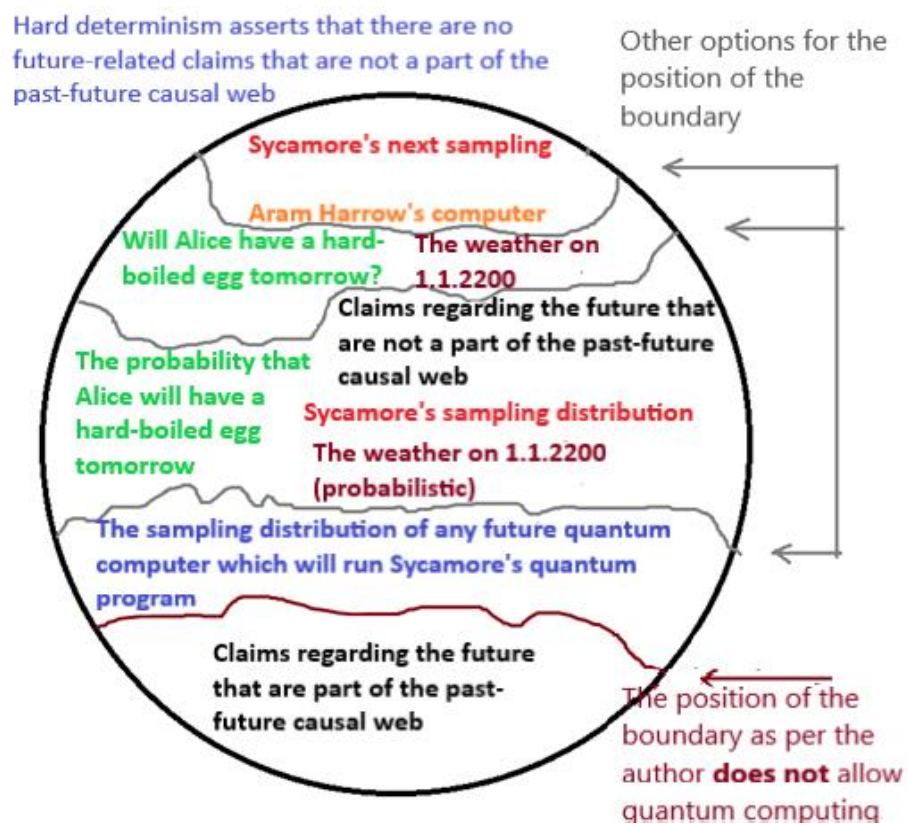

Figure 5: My theory regarding the infeasibility of quantum computers proposes a drastic option with far-reaching conclusions concerning the position of the boundary between those claims which are, in principle, predictable, and those that are not. It is possible to consider any other options for the position of the boundary: for instance,

one can assume that it would be possible to accurately describe Sycamore's sampling distribution. My theory asserts that this distribution would be chaotic and unpredictable; It is reasonable to think (and many, indeed, do) that it would be possible to build a quantum computer which performs the Sycamore sampling precisely as predicted (without noise), but my theory negates this as well.

## 10.     Summary

The purpose of the argument presented in this paper is to negate the perception that deterministic laws of nature preclude free will. I have shown that elements of the theory according to which quantum computers are not possible imply that Alice's future – much like the future of far simpler quantum systems – is not predictable: not only is it impossible to predict answers to simple questions regarding Alice's future as a matter of principle (Section 5), but it is also impossible to view answers to such questions as part of the past–future causal fabric (Section 7), and a correct definition of "Alice" points to the fact that it is impossible to attribute physical meaning in the present to claims concerning Alice's choices in the future (Section 8). That is, a physically meaningful definition of Alice in the present necessitates multiple options for her future and negates the claim that Alice's decisions in the present have already been determined by the past. It is thus possible that there is a certain component in Alice's future that depends solely on her decisions in the present. We emphasize that "unpredictability" refers also to probabilistic predictions, and that "multiple possibilities for the future" refers to a situation where there are multiple probabilistic possibilities for the future, rather than a single probability distribution describing it.

My explanation concurs with the approaches that view Alice herself as a higher-layer emergent concept, and its uniqueness lies in the fact that the principles of the emergent description apply to much broader physical systems, and that for these systems (including Alice) the emergent description has no alternative, either as a more basic or a more accurate physical description. More specifically, the impossibility of predicting the future (even if only probabilistically) also applies to systems much simpler than Alice, like the Sycamore computer, and the aforementioned move we discussed, from the epistemic argument for predictability to the ontological argument for multiple future possibilities, also applies in greater generality.

Considering free will as a real phenomenon is often associated with metaphysical beliefs, non-physical assumptions, and defiance of science. By contrast, central to my analysis is the position that an ontological description of objects in the physical world cannot be based on physically meaningless notions.

A successful resolution of the apparent contradiction between the laws of nature and free will would give strong support to the stance that free will is a real phenomenon. I would like to point out, however, that there are several philosophical approaches, starting with the perception that "everything is foreseen yet freedom of choice is granted,"[14] that do not consider even the full ability to predict Alice's decisions as an obstacle to the assertion that her will is free. On the other hand, even if we accept that "not everything is foreseen" and that free will is not rejected by the laws of nature, we should still ask whether there is free will at all, and if it does exist, what it actually means and what is its nature. This is a philosophical discussion that is closer to understanding humans than to understanding the laws of nature.[15]

The Hebrew University of Jerusalem and Reichman University, Herzliya

**Acknowledgements**

I would like to extend my gratitude to quite a few colleagues for their helpful and incisive comments on drafts of this article, and for fruitful conversations on the issue of free will. My research is supported by ERC grant 834735.

This manuscript is an extended English translation of the author's Hebrew essay "מחשבים קוונטיים, פרדיקטביליות ורצון חופשי".

---

[14] הכל צפוי והרשות נתונה, *Pirkei Avot* (Sayings of the Fathers), Chapter 3.
[15] See, Frankfurt (1969, 1971), Wolf (1987), Usher (2006, 2018), Gilboa (2009), Malkiel (2013), and Drai (2016).

30

## Appendix: Recent Experiments on NISQ Computers

I will briefly describe the experimental efforts to demonstrate quantum computational supremacy and to build stable qubits based on quantum error-correction. As previously mentioned, my theory negates the possibility that these goals can be achieved. The most dramatic progress has been made with regard to "quantum supremacy." In a 2019 paper published in *Nature*, a group from Google (Arute et al. 2019) claimed that quantum supremacy had been achieved by means of sampling performed in 300 seconds on the Sycamore 53-qubit quantum computer. The paper further claimed that the same sampling task would take a classical supercomputer 10,000 years to carry out. This achievement was compared by some to the Wright brothers' first flight, to Fermi's demonstration of a nuclear chain reaction, and to the discovery of the Higgs particle. In an article published in *Science* in 2020, a USTC group (Zhong et al. 2020) claimed that it had built another quantum system, based on photons, that achieved samples (again, within 300 seconds) that would take billions of computation years to carry out on a classical supercomputer.

In both cases classical algorithms that efficiently perform the computational task were later discovered, thereby largely refuting the quantum supremacy claims. For the Google experiment, researchers reduced the required running time by more than a billion times, from 10,000 years to several minutes (Pan, Chen, and Zhang 2021, Kalachev, Panteleev, and Yung 2021, and others). The quantum supremacy claim relating to the photonic system has, to a large extent, been refuted by a paper by Guy Kindler and myself (Kalai and Kindler 2014). This paper made use for the first time of the Fourier methods (specifically the Hermite–Fourier expansion) for the purpose of studying NISQ systems (specifically "boson sampling").

Some progress has been made by several groups with regard to the building of stable qubits based on quantum error-correcting codes by means of NISQ systems. However, this progress is not significant enough to be in tension with my theory. Another research direction, for achieving stable topological qubits, is still in its preliminary stages and is encountering many obstacles. (My theory also negates the possibility of achieving stable topological qubits.)

It is reasonable to think that within a few years the picture will clear up about the present experimental claims regarding NISQ computers, and specifically the claims of "quantum supremacy." A time scale of a few decades will be required to determine the possibility of quantum computing, while the ancient questions of predictability and free will undoubtedly occupy us much longer.

The experiments performed on the Sycamore computer provide an opportunity to examine several aspects of my theory. An analysis of samples of the 12-qubit quantum computer (Figure 6) demonstrates non-stationary behavior and also suggests chaotic behavior. The samples also confirm the exponential decay of Fourier coefficients caused by noise (Figure 7). (Both figures are part of a joint work with Yosef Rinott and Tomer Shoham.)

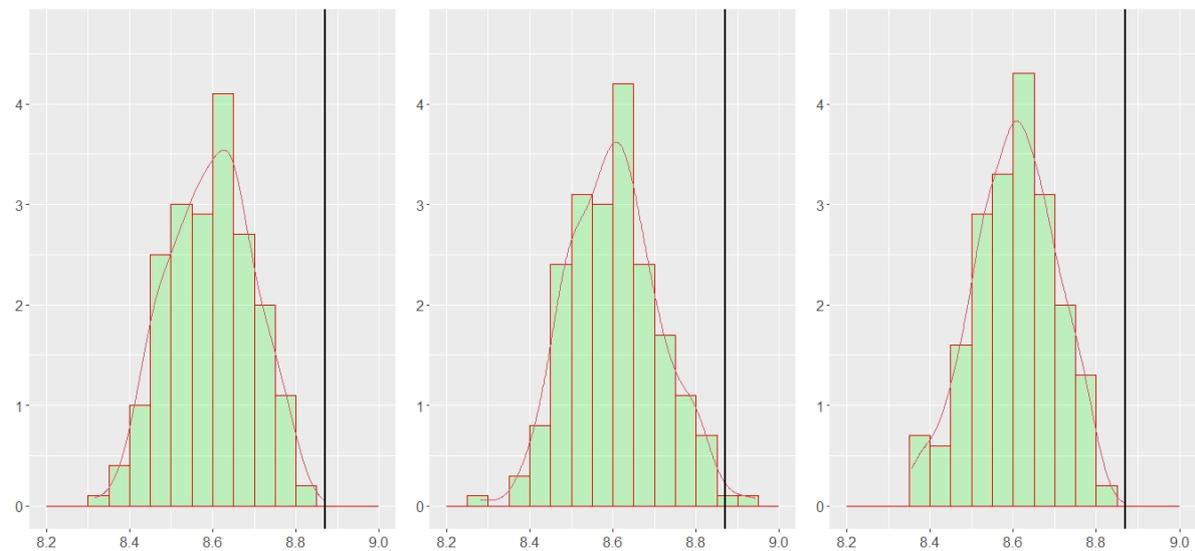

Figure 6: A comparison between the distribution of the first half as opposed to the second half of .5M samples supports the claim of a non-stationary and perhaps even chaotic behavior. The figure shows how the distance between the empirical distributions of the two halves of the sample (shown by the vertical line) is significantly larger than the distance between the two halves (as described by the histogram) when we randomly divide the sample into two equal parts. (The figure shows it for three random circuits.)

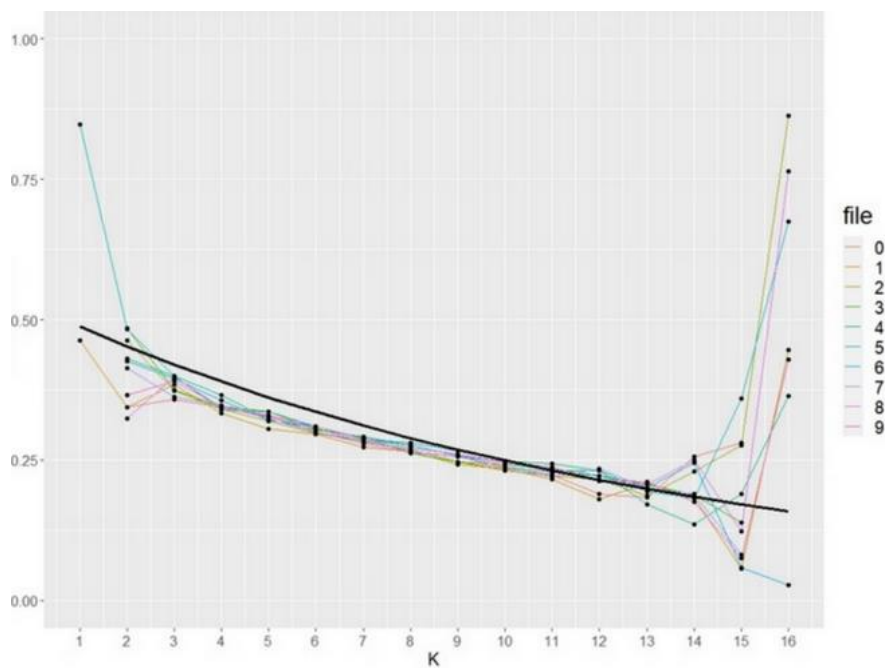

Figure 7: The decay of Fourier–Walsh contributions as predicted by the theory (bold black curve) and as demonstrated by samples of a 16-qubit Sycamore computer (ten random circuits).